\documentclass[preprint,12pt,authoryear]{elsarticle}
\usepackage{amssymb,amsfonts,amsthm}
\usepackage[fleqn]{amsmath}
\usepackage{geometry}
\usepackage{graphics,subfigure}
\usepackage{hyperref}
\usepackage{pifont}
\usepackage{natbib}
\usepackage{color}
\usepackage{bbm}
\usepackage{bm}
\usepackage{appendix}
\numberwithin{equation}{section}
\begin{document}

\begin{frontmatter}
\title{Bending-induced director reorientation in a nematic liquid crystal elastomer bonded to a hyperelastic substrate}

\author[add1,add2]{Yang Liu}
\author[add1]{Wendi Ma}
\author[add3]{Hui-Hui Dai\corref{cor1}}
\cortext[cor1]{corresponding author}
\ead{mahhdai@cityu.edu.hk}

\address[add1]{Department of Mechanics, Tianjin University, Tianjin 300350, China}
\address[add2]{Tianjin Key Laboratory of Modern Engineering Mechanics, Tianjin 300350, China}
\address[add3]{Department of Mathematics, City University of Hong Kong, 83 Tat Chee Avenue, Kowloon Tong, Hong Kong}

\begin{abstract}
In this paper, the two-dimensional pure bending of a hyperelastic substrate coated by a nematic liquid crystal elastomer (abbreviated as NLCE) is studied within the framework of nonlinear elasticity. The governing system, arising from the deformational momentum balance, the orientational momentum balance and the mechanical constraint, is formulated, and the corresponding exact solution is derived for a given constitutive model. It is found that there exist two different bending solutions. In order to determine which the preferred one is, we compare the total potential energy for both solutions and find that the two energy curves may have an intersection point at a critical value of the bending angle $\alpha_c$ for some material parameters. In particular, the director $\bm n$ abruptly rotates $\dfrac{\pi}{2}$ from one solution to another at $\alpha_c$, which indicates a director reorientation (or jump). Furthermore, the effects of different material and geometric parameters on the bending deformation and the transition angle $\alpha_c$ can be revealed using the obtained bending solutions. Meanwhile, the exact solution can offer a benchmark problem for validating the accuracy of approximated plate models for liquid crystal elastomers.
\end{abstract}

\end{frontmatter}

\section{Introduction}
Liquid crystal elastomers (LCEs) combine both the hyperelasticity of rubber-like solids and attractive features of liquid crystals \citep{book2007}. Among various LCEs, the nematic liquid crystal elastomer (NLCE) is the simplest one, for which the anisotropy can be measured by a director $\bm n$. NLCEs can be viewed as a kind of promising intelligent material due to its quick and controlled reaction in response to various stimuli, such as light \citep{apl2011,wy2012}, magnetic or electric field \citep{uht2005,uht2006,kw2009,wk2010}, and temperature variation \citep{sm2014,cai2017}. These desirable properties justify their applications to artificial muscles \citep{gennes1997,tian2018}, flexible robotics \citep{de2015,am2019}, soft actuators \citep{yu2016,sm2019},  etc. Interested readers are referred to the monograph by \cite{book2007} where an exhaustive introduction to the physics of NLCE and extensive references can be found.

Indeed, NLCEs can undergo a large elastic deformation as well. It is therefore of fundamental significance to model their mechanical features, especially plentiful nonlinear behaviors. Accounting for nematic order, an extension of rubber elasticity was proposed by \cite{bladon1993}. Later, \cite{arma2002} presented a modified energy form incorporating the quasi-convexity. In continuum-mechanical framework, the pioneering work was accomplished by \cite{fried1999}, where the strain-energy function is assumed to be a function of deformation gradient, orientation (or director), and orientation gradient. In doing so, besides the conventional equilibrium equations, the governing system contains an additional vector equation owing to orientational momentum balance. Furthermore,  attempts and progresses were also made to improve the existing models and approaches based on the neo-classical theory \citep{je2002,fs2004,de2009}, continuum theory \citep{chenfried2006}, or other rubber elasticities \citep{ad2012}. Recently, a new continuum theory was presented by \cite{huo2019} according to the dissipation principle.

In principle, most experimentally observed phenomena can be reproduced using the above mentioned models, for instance, soft elasticity \citep{pre1999, fk2002, huo2019}. In addition, some basic deformations for NLCEs can also be investigated theoretically such as inflation \citep{chenfried2006}, disclination \citep{fr2006}, and pure bending \citep{pence2006}. In particular, a finite bending can serve as a
benchmark problem for validating the accuracy of plate theories for NLCEs, because the exact solution is position dependent and can be solved analytically. It is noted that the investigation by \cite{pence2006} employed a constitutive model with a solid phase, a smectic phase and a liquid phase \citep{arma2002}. It seems that there is a lack of studies concerning finite bending of NLCEs using the continuum-mechanical theory proposed by \cite{fried1999}. Furthermore, we emphasize that an NLCE sample usually occupies a very thin thickness around several hundreds microns ($\mu$m) while the other two dimensions can be dozens of times greater than the thickness. Motivated by these two facts, the current work then focuses on the bending behavior of an NLCE film bonded to a hyperelastic material and aim at providing some new insights in a rational way. 

The rest of this paper is organized as follows. In Section 2, the governing system of an NLCE-substrate structure under finite bending is formulated. In Section 3, the exact solution is derived, and two different bending solutions with mutually orthogonal directors are found. Section 4 addresses the issue of the competition between two solutions by comparing their total potential energies. It turns out that, for some parameter choices, a director reorientation (or jump) can be triggered when the bending angle reaches a critical value. Finally, some conclusions are presented in Section 5.

\section{Problem formulation}
\begin{figure}
\centering\includegraphics[scale=0.7]{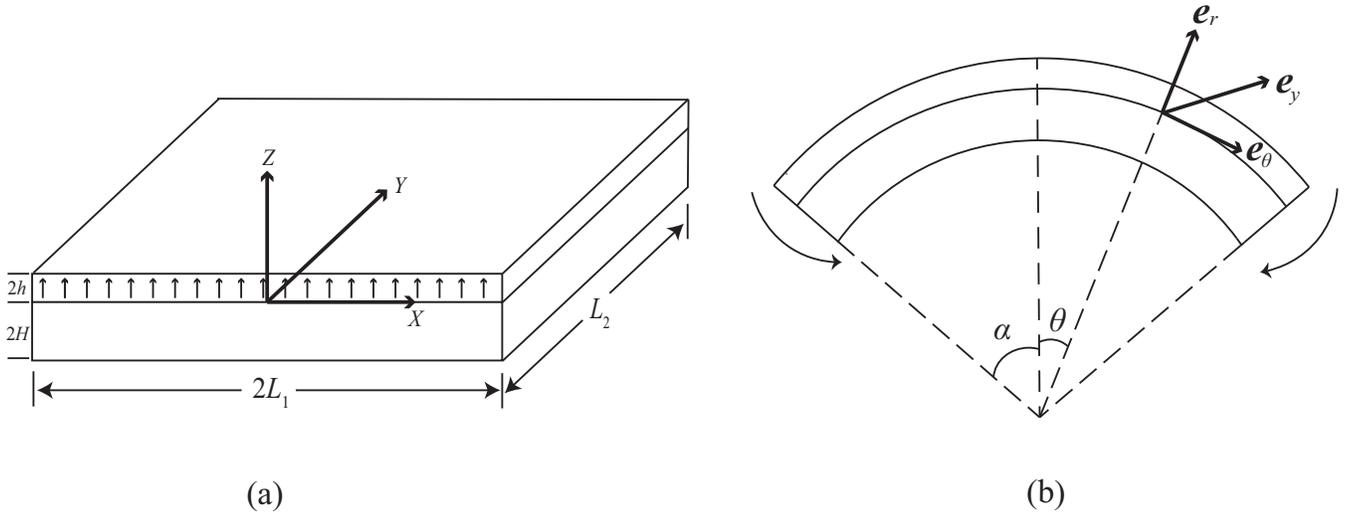}\caption{(a) A 3D illustration of an NLCE-substrate structure at the stress-free state; (b) The two-dimensional pure bending state induced by a bending moment.}\label{fig1}
\end{figure}

Consider a film composed of a nematic liquid crystal elastomer (NLCE) bonded to a hyperelastic substrate subjected to a bending angle/moment on the two edges. In the reference state, the thicknesses for the NLCE and substrate are given by $2h$ and $2H$, respectively. As shown in Figure \ref{fig1}, the Cartesian coordinate system $(X,Y,Z)$ is used in the stress-free state, and the width and length are represented by $2L_1$ and $L_2$. The origin is located at the center of the interface which is assumed to be perfectly bonded during the deformation. We suppose that the structure deforms into a sector of a cylindrical tube under a bending angle/moment. Therefore, the cylindrical polar coordinate system $(\theta, y, r)$ is employed in the current state, and the deformation is described by \citep{ogden1997}
\begin{equation} 
\begin{cases}
\theta=\dfrac{\alpha}{L_1}X,~~-L_1\leq X\leq L_1,\\
y=Y,~~-L_2\leq Y\leq L_2,\\
r=r(Z),~~-2H\leq Z\leq 2h,
\end{cases}\label{eq2_1}
\end{equation}
where $\alpha$ stands for the bending angle. In particular, we have specified a plane-strain deformation in $(X,Z)-$plane. Without loss of generality, we set $L_1=1$ in the subsequent analysis such that $h$ and $H$ indicate the thickness-width ratios from now on. 

Bearing in bind that an NLCE contains a director which can be used to measure the alignment of mesogens. In this study, as indicated in Figure \ref{fig1}, we choose that the referential director is along the $Z-$axis. Furthermore, it is assumed that the current director is also an in-plane vector. We hence get
\begin{align}
{\bm n}_0={\bm E}_3,~~{\bm n}=\textrm{cos}\beta{\bm e}_\theta+\textrm{sin}\beta{\bm e}_r,\label{eq2_2}
\end{align} 
where ${\bm n}_0$ is the initial director and $\beta$ is a constant that corresponds to the anti-clockwise angle between the ${\bm n}$ and ${\bm e}_\theta$. 

In order to distinguish different quantities for the NLCE film or substrate, we define a convention that a common notation with a subscript $f$ belongs to the NLCE film, or it belongs to the substrate with a subscript $s$. Moreover, if a quantity shares the same expression for the substrate and NLCE, we abandon any subscript. With respect to an orthonormal basis $({\bm E}_1,{\bm E}_2,{\bm E}_3)$ in the reference state and an orthonormal basis $({\bm e}_\theta,{\bm e}_y,{\bm e}_r)$ in the current state, the deformation gradient tensor for both layers of the current problem reads
\begin{align}
\mathbb{F}=\alpha r(Z){\bm e}_\theta\otimes{\bm E}_1+{\bm e}_y\otimes{\bm E}_2+ r'(Z){\bm e}_r\otimes{\bm E}_3,\label{eq2_3}
\end{align}
where the prime here and hereafter indicates the differentiation with respect to $Z$. 

Note that both the film and substrate consist of incompressible materials. For the NLCE film, we employ a continuum model proposed by \cite{fried1999} where the strain-energy function is assumed to be a function of the deformation gradient $\mathbb{F}$, the director $\bm n$, and the orientation gradient tensor $\mathbb{G}=\textrm{Grad}(\bm n)$. Note that the operator ``Grad'' is evaluated in the reference configuration. In particular, we specify a two-constant constitutive model as follows \citep{fried1999,fk2002}:
\begin{align}
\nonumber \phi_f=&~\frac{\mu_f}{2}\bigg(\textrm{tr}(\mathbb{F}\mathbb{F}^{\textrm{T}})-\frac{s-1}{s}\mathbb{F}^{\textrm{T}}{\bm n}\cdot\mathbb{F}^{\textrm{T}}{\bm n}+(s-1)\mathbb{F}{\bm n}_0\cdot\mathbb{F}{\bm n}_0-\frac{(s-1)^2}{s}(\mathbb{F}^{\textrm{T}}{\bm n}\cdot {\bm n}_0)^2-3\bigg)\\&+\frac{\kappa(s-1)^2}{2s}\textrm{tr}(\mathbb{F}^{\textrm{T}}\mathbb{G}\mathbb{G}^{\textrm{T}}\mathbb{F}),\label{eq2_4}
\end{align}
where the superscript T stands for the transpose. The last term can be regarded as a gradient-energy density and the previous terms is consistent with the neo-classical free energy proposed by \cite{bladon1993}. In the above formulation, the positive constants $\mu_f$ and $\kappa$ signify the shear modulus and Frank constant, respectively, and $s$ is the step-length anisotropy. We emphasize that $\kappa$ originally has the same dimension of force. Yet the setting $L_1=1$ implies that $\kappa$ has been scaled by $L_1$ such that it immediately has the dimension of 2D stress. In reality, the parameters $s$ is larger than zero. When $s$ equals to unity, (\ref{eq2_4}) reduces to the classical neo-Hookean model, or otherwise, $s<1$ or $s>1$ represent that the molecules are oblate or prolate, respectively. For the hyperelastic substrate, we adopt the incompressible neo-Hookean material with the following strain energy function
\begin{align}
\phi_s=\frac{\mu_s}{2}\bigg(\textrm{tr}(\mathbb{F}\mathbb{F}^{\textrm{T}})-3\bigg),\label{eq2_5}
\end{align}
with $\mu_s$ the corresponding shear modulus.

In general, the equilibrium equations for an NLCE in the current framework arise from the deformational momentum balance and orientational momentum balance. Meanwhile, the convention that the director always be a unit vector gives rise to two additional constraints on the director and orientation gradient, apart from the incompressibility constraint. Consequently, more Lagrange multipliers are involved in the governing system. However, it has been clarified by \cite{fried1999} that the multipliers in response to the constraint $\bm n\cdot\bm n=1$, which can be eliminated from the governing system, are of negligible importance. In the subsequent analysis, we follow a technique in \cite{chenfried2006} to get rid of them. 

For the particular form (\ref{eq2_4}), the nominal stress tensor for the NLCE is given by
\begin{align}
\nonumber\mathbb{S}_f(\mathbb{F},\bm n,\mathbb{G})=&~\mu_f\left(\mathbb{F}^\textrm{T}-\dfrac{s-1}{s}\mathbb{F}^\textrm{T}{\bm n}\otimes{\bm n}+(s-1){\bm n}_0\otimes\mathbb{F}{\bm n}_0-\dfrac{(s-1)^2}{s}(\mathbb{F}^\textrm{T}{\bm n}\cdot{\bm n}_0){\bm n}_0\otimes{\bm n}\right)\\&+\dfrac{\kappa(s-1)^2}{s}\mathbb{F}^\textrm{T}\mathbb{G}\mathbb{G}^\textrm{T}-p_f(Z)\mathbb{F}^{-1},\label{eq2_6}
\end{align}
where $p_f(Z)$ is a pressure associated with the incompressibility constraint, while the modified internal orientational force $\bm \pi$ and the modified orientational stress tensor $\mathbb{T}$ read
\begin{align}
&{\bm \pi}(\mathbb{F},\bm n,\mathbb{G})=-\dfrac{\mu_f(s-1)}{s}\left(\mathbb{F}\mathbb{F}^\textrm{T}+(s-1)\mathbb{F}{\bm n_0}\otimes\mathbb{F}{\bm n_0}\right){\bm n},\label{eq2_7}\\
&\mathbb{T}(\mathbb{F},\bm n,\mathbb{G})=\dfrac{\kappa(s-1)^2}{s}\mathbb{G}^\textrm{T}\mathbb{F}\mathbb{F}^\textrm{T}.\label{eq2_8}
\end{align}
Accordingly, in light of (\ref{eq2_5}), the nominal stress tensor for the substrate takes the form
\begin{align}
\mathbb{S}_s(\mathbb{F})=\mu_s\mathbb{F}^\textrm{T}-p_s(Z)\mathbb{F}^{-1},\label{eq2_9}
\end{align}
with $p_s(Z)$ the corresponding pressure for the substrate.

Next, we calculate the orientation gradient tensor $\mathbb{G}$ according to (\ref{eq2_2}) to obtain
\begin{align}
\mathbb{G}=\alpha\textrm{sin}\beta{\bm e}_\theta\otimes{\bm E}_1-\alpha\textrm{cos}\beta{\bm e}_r\otimes{\bm E}_1.\label{eq2_10}
\end{align}
Since both the NLCE film and the substrate are composed of incompressible materials, we have the following constraint equation
\begin{align}
\textrm{Det}(\mathbb{F})=\alpha r(Z)r'(Z)=1.\label{eq2_11}
\end{align}

Neglecting the all body forces, the governing system for an NLCE-substrate structure for a static problem gives
\begin{align}
\nonumber &\textrm{Div}\mathbb{S}_f={\bm 0}, ~~\textrm{in}~~\left[-L_1,L_1\right]\times\left[-L_2,L_2\right]\times\left[0,2h\right],\\&\nonumber
\left(\mathbb{I}-{\bm n}\otimes{\bm n}\right)\left(\textrm{Div}\mathbb{T}+\bm{\pi}\right)={\bm 0},~~\textrm{in}~~\left[-L_1,L_1\right]\times\left[-L_2,L_2\right]\times\left[0,2h\right],\\&\textrm{Div}\mathbb{S}_s={\bm 0}, ~~\textrm{in}~~\left[-L_1,L_1\right]\times\left[-L_2,L_2\right]\times\left[-2H,0\right].\label{eq2_12}
\end{align}
We mention that $(\ref{eq2_12})_2$ is the orientational momentum balance equation and the factor $\left(\mathbb{I}-{\bm n}\otimes{\bm n}\right)$ is used to eliminate the Lagrange multipliers that react to the constraint $\bm n\cdot\bm n=1$.

The upper and lower surfaces are granted to be traction-free, which leads to the following boundary conditions:
\begin{align}
\mathbb{S}_f^\textrm{T}{\bm E}_3\big|_{Z=2h}={\bm 0},~~\mathbb{S}_s^\textrm{T}{\bm E}_3\big|_{Z=-2H}={\bm 0},~~\left(\mathbb{I}-{\bm n}\otimes{\bm n}\right)\mathbb{T}^\textrm{T}{\bm E}_3\big|_{Z=2h,0}={\bm 0}.\label{eq2_13}
\end{align}
The last formula in (\ref{eq2_13}) arises from the natural boundary condition of the orientational stress tensor in the variational problem. 

As mentioned earlier, the interface between two layers keeps perfectly bonded during the deformation. Thus, both the displacement and traction should be continuous on the interface. Remember that the displacement continuity condition is satisfied automatically according to the prescribed bending deformation $(\ref{eq2_1})_3$, therefore the traction continuity condition furnishes
\begin{align}
\mathbb{S}_s^\textrm{T}{\bm E}_3\big|_{Z=0}=\mathbb{S}_f^\textrm{T}{\bm E}_3\big|_{Z=0}.\label{eq2_14}
\end{align}

Currently, the governing system together with all required boundary conditions and continuity conditions are established. It can be seen that there are in total three unknowns $r(Z)$, $p_f(Z)$, and $p_s(Z)$. Once those unknowns are solved, one can find the relation between the bending moment $M$ on the two edges and the bending angle $\alpha$:
\begin{align}
M=L_2\left(\int_{-2H}^0r(Z)S_{s11}\textrm{d}Z+\int_0^{2h}r(Z)S_{f11}\textrm{d}Z\right),
\end{align}
and then one of them can be determined if another one is prescribed.

\section{Exact solutions}
In this section, we shall derive the exact solutions of $r(Z)$, $p_f(Z)$, and $p_s(Z)$ by solving the governing system (\ref{eq2_12}) with the boundary conditions (\ref{eq2_13}) and continuity condition (\ref{eq2_14}). First of all, on substituting (\ref{eq2_2}), (\ref{eq2_3}) and (\ref{eq2_10}) into (\ref{eq2_6})--(\ref{eq2_9}), we acquire
\begin{align}
\nonumber\mathbb{S}_f=&\left(\mu_f\alpha r(Z)-\dfrac{p_f(Z)}{\alpha r(Z)}-\dfrac{\mu_f(s-1)}{s}\alpha r(Z)\textrm{cos}^2\beta+\dfrac{\kappa(s-1)^2}{s}\alpha^3r(Z)\textrm{sin}^2\beta\right){\bm E}_1\otimes{\bm e}_\theta\\&\nonumber+\left(-\dfrac{\mu_f(s-1)}{s}\alpha+\dfrac{\kappa(s-1)^2}{s}\alpha^3\right)r(Z)\textrm{sin}\beta\textrm{cos}\beta {\bm E}_1\otimes{\bm e}_r+\bigg(\mu_f-p_f(Z)\bigg){\bm E}_2\otimes{\bm e}_y\\\nonumber&+\left(\dfrac{\alpha ^2 \kappa(s-1)}{s} -\alpha ^2 \kappa(s-1)-\frac{\mu_f(s-1) }{2}\right)r'(Z) \sin\beta\cos\beta {\bm E}_3\otimes{\bm e}_\theta\\\nonumber&+\bigg(-\frac{p(Z)}{r'(Z)}+\mu_f s r'(Z)+\left(\alpha^2\kappa(s-1)-\dfrac{\alpha^2\kappa(s-1)}{s}\right)r'(Z)\cos^2\beta\\&+\mu_f(1-s)r'(Z)\sin^2\beta \bigg){\bm E}_3\otimes{\bm e}_r,\label{eq3_1}
\end{align}
\begin{align}
{\bm \pi}=-\dfrac{\mu_f(s-1)}{s}\left(\alpha^2 r^2(Z)\textrm{cos}\beta{\bm e}_\theta+sr'(Z)^2\textrm{sin}\beta{\bm e}_r\right),\label{eq3_2}
\end{align}
\begin{align}
\mathbb{T}=\dfrac{\kappa(s-1)^2}{s}\left(\alpha^3r^2(Z)\textrm{sin}\beta{\bm E}_1\otimes{\bm e}_\theta-\alpha r'^2(Z)\textrm{cos}\beta{\bm E}_1\otimes{\bm e}_r\right).\label{eq3_3}
\end{align}
\begin{align}
\mathbb{S}_s=\left(\mu_s\alpha r(Z)-\dfrac{p_s(Z)}{\alpha r(Z)}\right){\bm E}_1\otimes{\bm e}_\theta+\left(\mu_s-p_s(Z)\right){\bm E}_2\otimes{\bm e}_y+\left(\mu_s r'(Z)-\dfrac{p_s(Z)}{r'(Z)}\right){\bm E}_1\otimes{\bm e}_\theta.\label{eq3_4}
\end{align}

Solving the incompressibility condition (\ref{eq2_11}) dircetly yields 
\begin{align}
\left(r^2(Z)\right)'=\frac{2}{\alpha}~~\Rightarrow~~r(Z)=\sqrt{\dfrac{2Z}{\alpha}+C_1},\label{eq3_5}
\end{align}
where $C_1$ is an integration constant to be determined by using the boundary conditions and continuity condition. Next,  
it follows from $(\ref{eq2_12})_3$ and (\ref{eq3_4}) that 
\begin{align}
p_s(Z)=\dfrac{\mu_s}{2}\left(\dfrac{1}{C_1\alpha^2+2\alpha Z}-(C_1\alpha^2+2\alpha Z)\right)+C_2,\label{eq3_6}
\end{align}
with $C_2$ another integration constant. For the NLCE film, substituting the expression (\ref{eq3_1}) into equation $(\ref{eq2_12})_1$ furnishes the following two equations:
\begin{align}
&\dfrac{s-1}{s} \sin\beta\cos\beta \bigg(r''(Z) \left(\alpha ^2 \kappa (s-1)+s\mu_f\right)+\alpha ^2 r(Z) \left(\alpha ^2 k (s-1)+\mu_f \right) \bigg)=0,\label{eq3_7}
\\\nonumber&p_f(Z) \left(\frac{r''(Z)}{r'(Z)^2}+\frac{1}{r(Z)}\right)+\dfrac{1}{2s}\bigg(\left(s((s-1) \cos2 \beta+s+1)\mu _f +2 \kappa(s-1)^2 \alpha ^2 \cos^2\beta\right)r''(Z)\\&-2 \alpha ^2\left(\left(s-(s-1) \cos ^2\beta\right)\mu _f +\alpha ^2 \kappa  (s-1)^2 \sin ^2\beta\right)r(Z)\bigg)-\frac{p_f'(Z)}{r'(Z)}=0.\label{eq3_8}
\end{align}
Furthermore, specializing equation $(\ref{eq2_12})_2$ according to the expressions (\ref{eq3_2}) and (\ref{eq3_3}) for ${\bm \pi}$ and $\mathbb{T}$, we find
\begin{align}
\dfrac{(s-1)\sin \beta\cos\beta}{s}\bigg(\alpha ^2 r(Z)^2 \left(\alpha ^2 \kappa (s -1)+\mu_f \right)-r'(Z)^2 \left(\alpha ^2 \kappa (s-1)+s\mu_f \right)\bigg){\bm n}^{\bot}=0,\label{eq3_9}
\end{align}
where the vector ${\bm n}^{\bot}=\sin \beta {\bm e}_\theta-\cos\beta{\bm e}_r$ is perpendicular to ${\bm n}$. Remember that $r(Z)$ is specified by (\ref{eq3_5}), equations (\ref{eq3_7}) and (\ref{eq3_9}) hold if and only if $\sin\beta\cos\beta=0$, which is equivalent to 
\begin{align}
\beta=\dfrac{\pi}{2}~~\textrm{or}~~\beta=0.\label{eq3_10}
\end{align}
Indeed, for $\beta=\dfrac{\pi}{2}$, (\ref{eq2_2}) generates ${\bm n}={\bm e}_r$ while (\ref{eq2_2}) yields ${\bm n}={\bm e}_\theta$ for $\beta=0$. Moreover, the last boundary condition in $(\ref{eq2_13})_3$ is satisfied automatically when applying  (\ref{eq3_10}).

At present, we obtain two equilibrium solutions of $\beta$ given in (\ref{eq3_10}). It can be seen that the two solutions are disconnected branches. We mention that the similar phenomenon was found by \cite{chenfried2006} when considering the inflation of an NLCE tube. Generally, at least one of these may be unstable in the sense that it corresponds to a higher total potential energy. Next, the two solutions will be studied separately.

$\textbf{Case I}:~\beta=\dfrac{\pi}{2}$

In the case of $\beta=\dfrac{\pi}{2}$, equation (\ref{eq3_8}) reduces to
\begin{align}
p_f(Z) \left(\frac{r''(Z)}{r'(Z)^2}+\frac{1}{r(Z)}\right)-\frac{p_f'(Z)}{r'(Z)}+\mu _f r''(Z)-\frac{r(Z) \left(s\mu _f\alpha^2+(s-1)^2\kappa  \alpha ^4\right)}{s}=0.\label{eq3_11}
\end{align}
It can be seen that equation (\ref{eq3_11}) is just a first-order ordinary differential equation for $p_f(Z)$. Then we solve this equation to obtain
\begin{align}
p_f(Z)=-\dfrac{\kappa Z \alpha^3(s-1)^2}{s}+\mu\left(\dfrac{1}{\alpha(4Z+2\alpha C_1)}-\alpha Z-\dfrac{1}{2}\alpha^2C_1\right)+C_3,\label{eq3_12}
\end{align}
where $C_3$ is the third integration constant. 

Bearing in mind that the boundary conditions and continuity condition are still left. We then substitute (\ref{eq3_5}), (\ref{eq3_6}) and (\ref{eq3_12}) into $(\ref{eq2_13})_2$ and (\ref{eq2_14}) to offer two algebraic equations. By solving them,  $C_2$ and $C_3$ can be expressed in terms of $C_1$ as follows:
\begin{align}
&C_2=\dfrac{\left(1+16\gamma^2h^2\alpha^2-8\gamma hC_1\alpha^3+C_1^2\alpha^4\right)\mu_s}{2\alpha(C_1\alpha-4\gamma h)},\label{eq3_13}\\&
C_3=\left(\dfrac{1}{2C_1\alpha^2}+\dfrac{C_1\alpha^2}{2}\right)\mu_f+\left(\dfrac{1-C_1^2\alpha^4+4\gamma hC_1\alpha^3}{(C_1\alpha-4\gamma h)C_1\alpha^2}\right)2\gamma h\mu_s,\label{eq3_14}
\end{align}
where $\gamma=H/h$ is a new parameter denoting the thickness ratio of the substrate and NLCE film. Finally, by use of (\ref{eq3_13}) and (\ref{eq3_14}), the traction-free boundary condition on the upper surface $(\ref{eq2_13})_1$ offers a cubic equation given by 
\begin{align}
\nonumber&\bigg(s \mu _f+ (s-1)^2 \kappa\alpha ^2+\gamma  s \mu _s\bigg)\alpha ^5 C_1^3-4  (\gamma -1) h\alpha ^4\bigg(s \mu _f+\gamma  s \mu _s+(s-1)^2 \kappa\alpha ^2\bigg)C_1^2\\&-\bigg( \left(16  \gamma  h^2 s\alpha ^2+s\right)\mu _f+16 \gamma  h^2(s-1)^2 \kappa\alpha ^4 +\gamma\left(16\gamma  h^2 s \alpha ^2+s\right)\mu _s\bigg)\alpha C_1+4 \gamma  h s \left(\mu _f-\mu _s\right)=0.\label{eq3_15}
\end{align}
Once all material and geometric parameters are specified, equation (\ref{eq3_15}) can identify a rational solution of $C_1$, and further all unknowns $r(Z)$, $p_s(Z)$, and $p_f(Z)$ can be determined according to (\ref{eq3_5}), (\ref{eq3_6}), (\ref{eq3_12}), respectively.

$\textbf{Case II}:~\beta=0$

Here we turn to the other solution $\beta=0$. In this case, we add a bar to all unknowns for the purpose of discrimination. Therefore, equation (\ref{eq3_8}) becomes 
\begin{align}
\bar{p}_f(Z) \left(\frac{\bar{r}''(Z)}{\bar{r}'(Z)^2}+\frac{1}{\bar{r}(Z)}\right)-\frac{\bar{p}_f'(Z)}{\bar{r}'(Z)}+\frac{\bar{r}''(Z) \left(s^2\mu _f+(s-1)^2\kappa\alpha ^2\right)- \mu _f \alpha ^2\bar{r}(Z)}{s}=0,\label{eq3_16}
\end{align}
which yields
\begin{align}
\bar{p}_f(Z)=\dfrac{(s-1)^2\kappa\alpha^2+s\mu^2}{2s(2Z+\bar{C}_1\alpha)\alpha}-\dfrac{(2Z+\bar{C}_1\alpha)\mu\alpha}{2s}+\bar{C}_3,
\end{align}
with $\bar{C}_1$ and $\bar{C}_3$ two unknown constants.

Likewise, from the boundary conditions and continuity condition, we find out the relations among $\bar{C}_1$, $\bar{C}_2$, $\bar{C}_3$:
\begin{align}
&\bar{C}_2=\dfrac{\left(1+16\gamma^2h^2\alpha^2-8\gamma h\bar{C}_1\alpha^3+\bar{C}_1^2\alpha^4\right)\mu_s}{2\alpha(\bar{C}_1\alpha-4\gamma h)},\\&
\bar{C}_3=\dfrac{(s-1)^2\kappa}{2s\bar{C}_1}+\left(\dfrac{\bar{C}_1\alpha^2}{2s}+\dfrac{s}{2\bar{C}_1\alpha^2}\right)\mu_f+\left(\dfrac{2\gamma h(1+4\gamma h\bar{C}_1\alpha^3-\bar{C}_1^2\alpha^4)}{\bar{C}_1(\bar{C}_1\alpha-4\gamma h)\alpha^2}\right)\mu_s,
\end{align}
and further a cubic equation for $\bar{C}_1$, which writes
\begin{align}
\nonumber &\bigg(\mu _f+\gamma  s \mu _s\bigg)\alpha ^5 \bar{C}_1^3-4 (\gamma -1) h \left(\mu _f+\gamma  s \mu _s\right)\alpha ^4 \bar{C}_1^2-\bigg(\left(16 \gamma  h^2\alpha ^2+s^2\right)\mu _f +(s-1)^2 \kappa \alpha ^2\\&+\gamma  s  \left(16\gamma  h^2\alpha ^2+1\right)\mu _s\bigg)\alpha  \bar{C}_1+4 \gamma  h \left(s^2 \mu _f-s \mu _s+ (s-1)^2\kappa \alpha ^2 \right)=0.
\end{align}

\begin{figure}[!htbp]
\centering\includegraphics[scale=1.2]{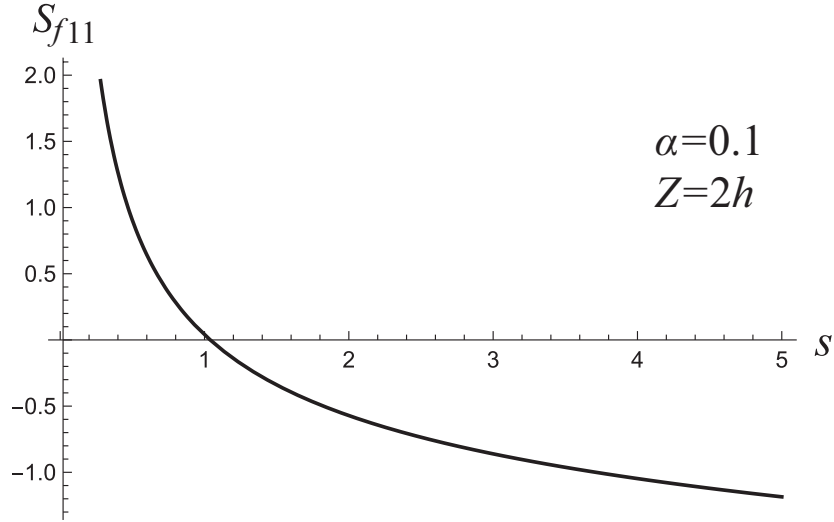}\caption{The relation between $S_{f11}$ and $s$ when $\alpha=0.1$ and $Z=2h$. Parameters are given by $h=0.05$, $\mu_f=1$, $\kappa=1$, $\gamma=1$.}\label{fig2}
\end{figure}

So far, the exact bending solutions for $\beta=\dfrac{\pi}{2}$ and $\beta=0$ are obtained. We are in a position to pick out the preferred one. In the next section, we intend to compare the total potential energies for both solutions for given parameters. To reduce the number of free parameters, we scale all quantities of stress dimension (2D stress in a plane-strain problem) by $\mu_s$. Thus, $\mu_f=2$ means that $\mu_f$ is twice larger than $\mu_s$. However, before moving forward, we should define a possibly rational parametric region. It can be seen in (\ref{eq2_4}) that there are in total three free parameters. Referring to \cite{fried1999,fk2002}, the constitutive model in (\ref{eq2_4}) has barely restrictions on those parameters except $\mu_f>0$, $s>0$, and $\kappa>0$. Nevertheless, carefully calculations reflect a fact that, as least for the current bending problem, the step-length anisotropy parameter $s$ cannot be an arbitrarily positive number. Figure \ref{fig2} plots the normal stress at two ends on the upper surface which suffers the maximum tensile deformation. But it is quite strange that the normal stress starts to be negative when $s$ is slightly greater than unity, which is contradicted with the practical deformation. We speculate that this may be relevant to the deviation of the strong ellipticity condition. Consequently, we restrict $s$ to be less than unity in the forthcoming illustrative examples.

\section{Critical angle to director reorientation} 
Since the present study focuses on the plane-strain bending, we therefore define the 2D total potential energy as
\begin{align}
\psi=\int_{-1}^1\int_0^{2h}\phi_f\textrm{d}Z\textrm{d}X+\int_{-1}^1\int_{-2H}^{0}\phi_s\textrm{d}Z\textrm{d}X,
\label{eq4_1}
\end{align}
which pertains to an experiment where the bending angle is controlled. If the bending moment is controlled in an experiment, an additional contribution of external load will join in (\ref{eq4_1}).

\begin{figure}[!htbp]
\centering
{\subfigure[$h=0.05$, $s=0.8$, $\mu_f=3$, $\kappa=1$, $\gamma=1$.]{\includegraphics[scale=0.9]{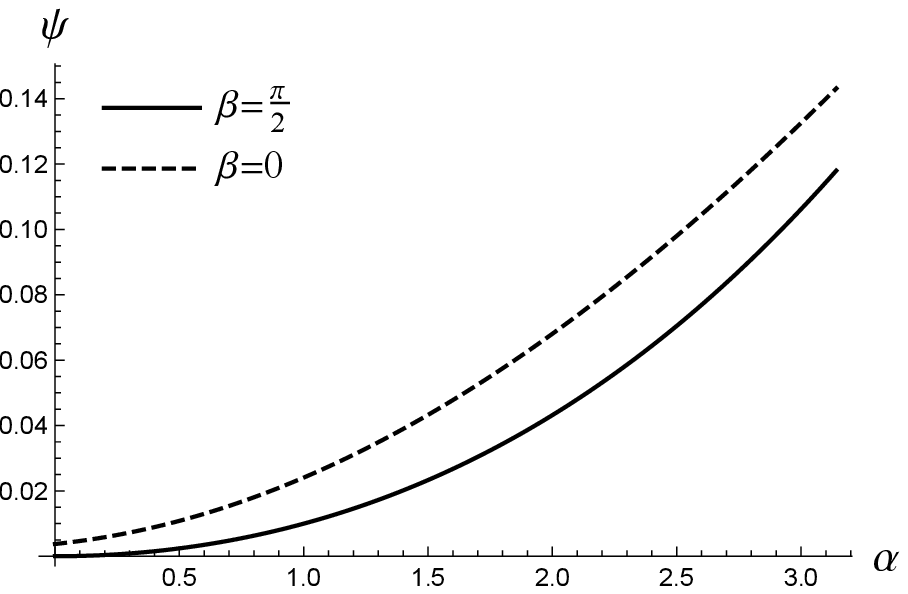}{\label{fig3a}}}\hspace{2mm}
\subfigure[$h=0.05$, $s=0.8$, $\mu_f=1$, $\kappa=1$, $\gamma=1$.]{\includegraphics[scale=0.9]{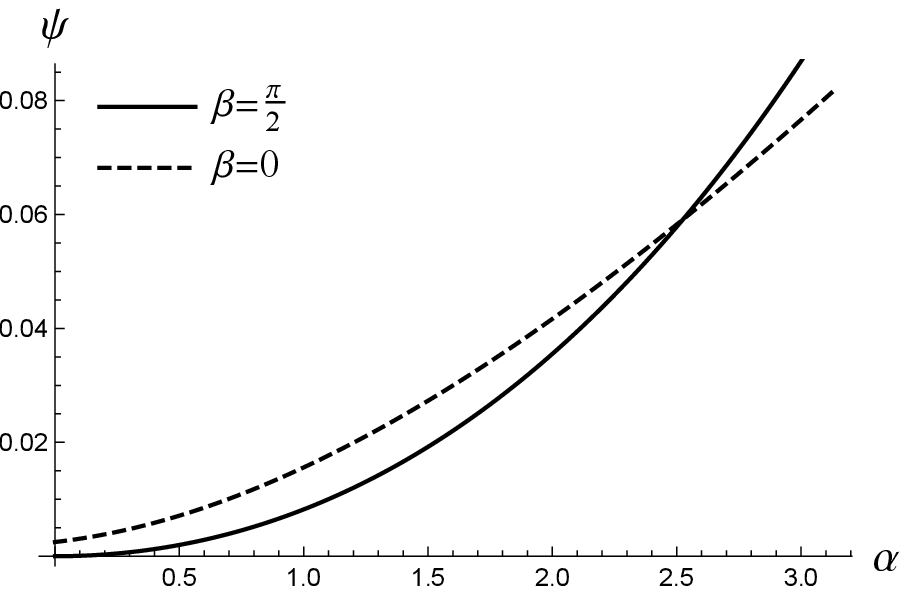}{\label{fig3b}}}}
\caption{Dependence of the energy $\psi$ on $\alpha$.}\label{fig3}
\end{figure}

Inserting all required quantities in Section 3 into (\ref{eq4_1}), the dependences of $\psi$ and the bending angle $\alpha$ for $\beta=\dfrac{\pi}{2}$ and $\beta=0$ can be characterized, respectively. Note that all calculations are performed in the help of $Mathematica$ \citep{math2019}. On assigning $h=0.05$, $s=0.8$, $\kappa=1$, and $\gamma=1$, we display the energy curves for $\mu_f=3$ and $\mu_f=1$ in Figure \ref{fig3}. It is observed that the total potential energy for $\beta=\dfrac{\pi}{2}$ is below than that for $\beta=0$ if the bending angle $\alpha$ is small. With $\alpha$ increased, the curve for $\beta=\dfrac{\pi}{2}$ is always lower in Figure \ref{fig3a}.  However, the two curves can intersect at $\alpha=2.525$ in Figure \ref{fig3b}. It can therefore be concluded that the director may rotate $\dfrac{\pi}{2}$ at a critical angle for certain parameter choice. We then define this transition value as $\alpha_c$. On the experimental side, \cite{prl1993} reported the director reorientation induced by uni-axial extension occurs if the axial stretch reaches around $1.13$. On the theoretical side, \cite{wz2013} was concerned with orientational transition in nematic gels, and they theoretically found the existence of director reorientation when the gels is under extension. Furthermore, \cite{chenfried2006} discovered the director jump in an inflated NLCE tube by employing the same model as in the present study. Although the finite bending of liquid crystal elastomers was investigated by \cite{pence2006} using a different material model, it seems that such an interesting phenomenon was not reported there. 
\begin{figure}[!htbp]
\centering
{\subfigure[$\kappa=1$, $s=0.8$, $h=0.05$, $\gamma=1$.]{\includegraphics[scale=0.9]{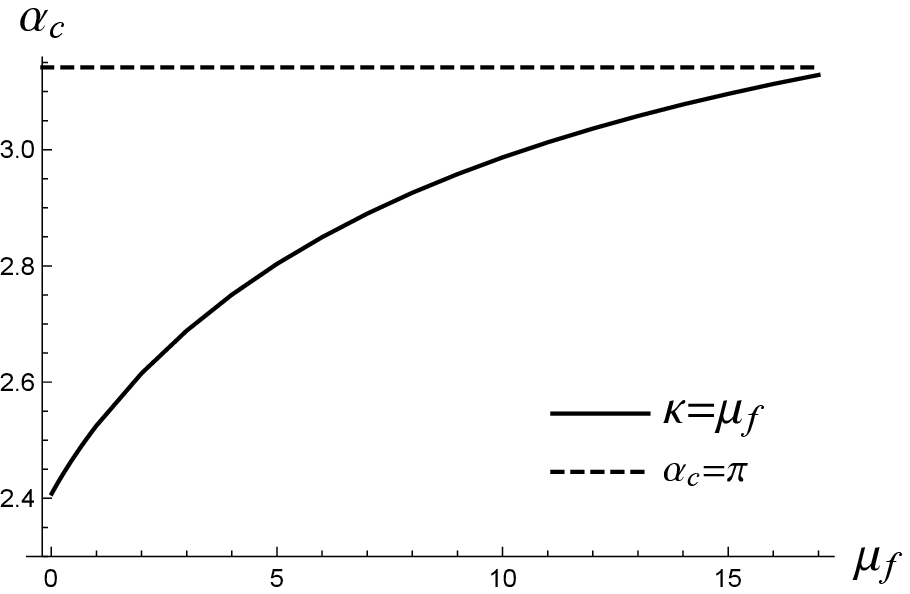}{\label{fig4a}}}\hspace{2mm}
\subfigure[$s=0.8$, $h=0.05$, $\gamma=1$.]{\includegraphics[scale=0.9]{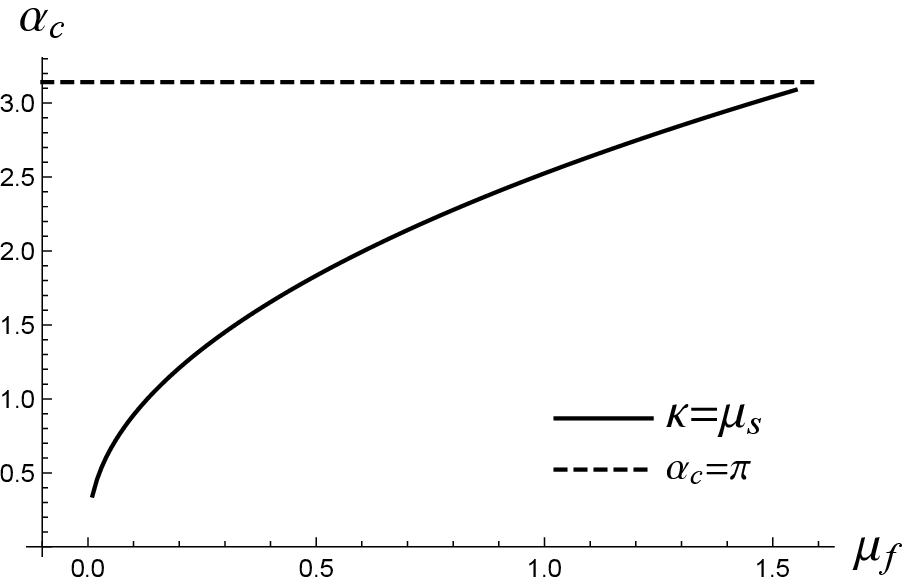}{\label{fig4b}}}
\subfigure[$\mu_f=1$, $s=0.8$, $h=0.05$, $\gamma=1$.]{\includegraphics[scale=0.9]{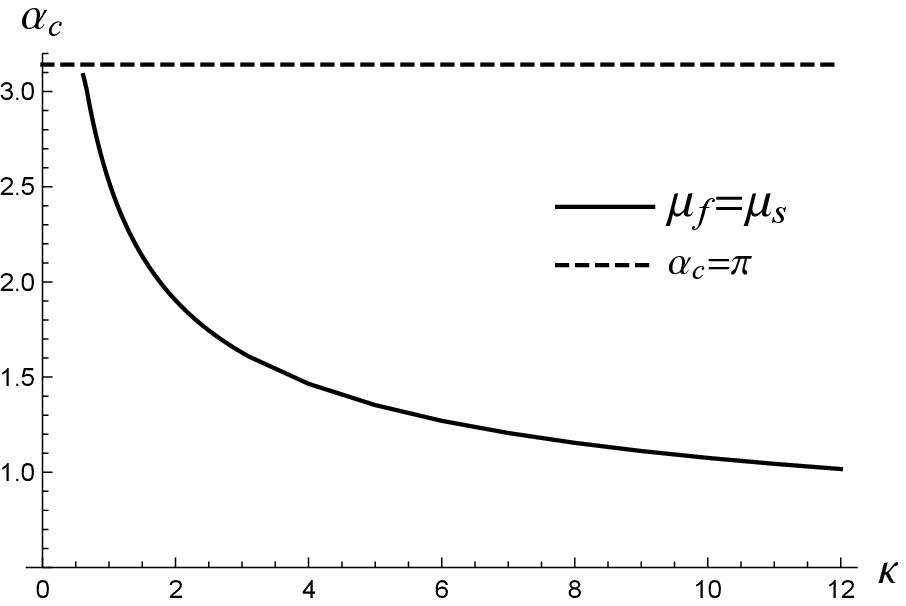}{\label{fig4c}}}
\subfigure[$\kappa=5$, $\mu_f=1$, $s=0.8$, $\gamma=1$.]{\includegraphics[scale=0.9]{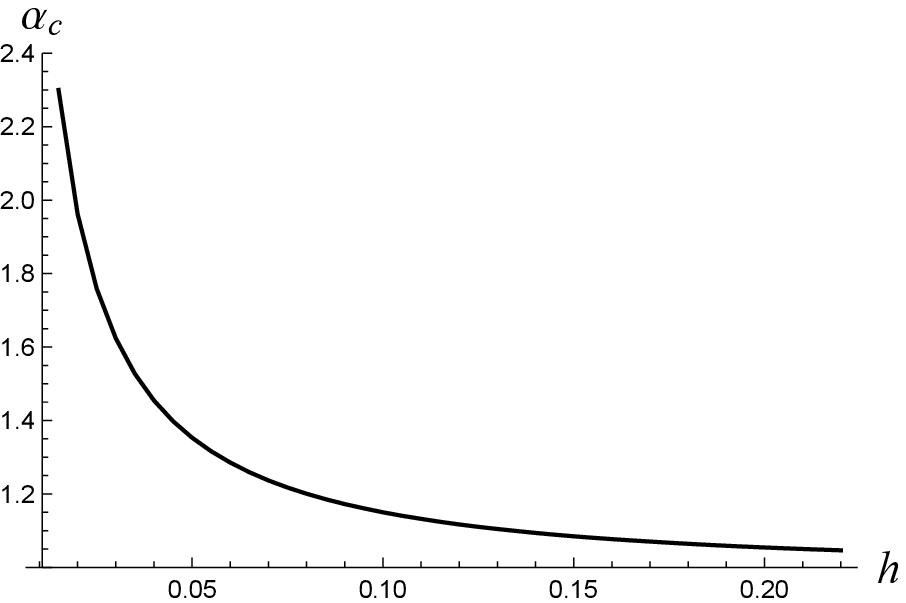}{\label{fig4d}}}
\subfigure[$\kappa=5$, $\mu_f=1$, $s=0.8$, $h=0.03$.]{\includegraphics[scale=0.9]{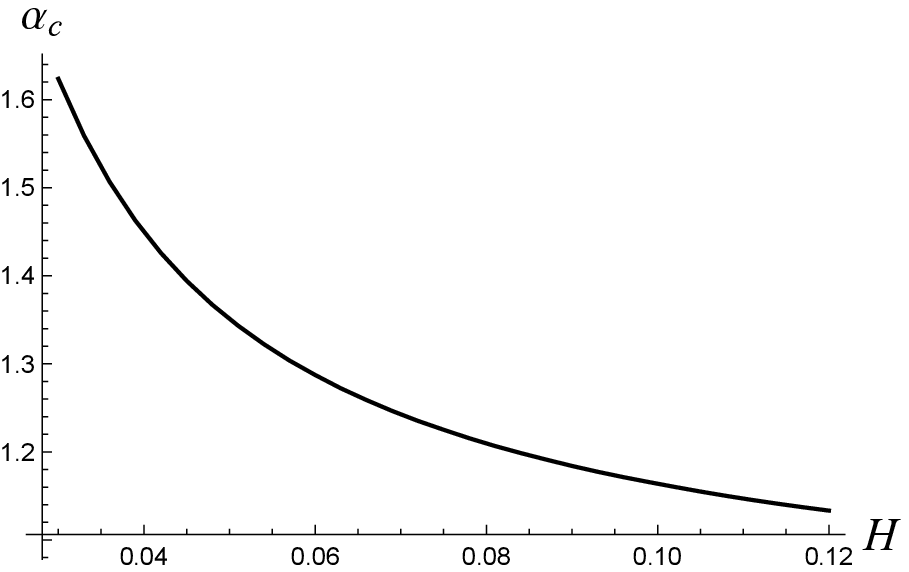}{\label{fig4e}}}
\subfigure[$\kappa=5$, $\mu_f=1$, $s=0.8$, $H=0.1$.]{\includegraphics[scale=0.9]{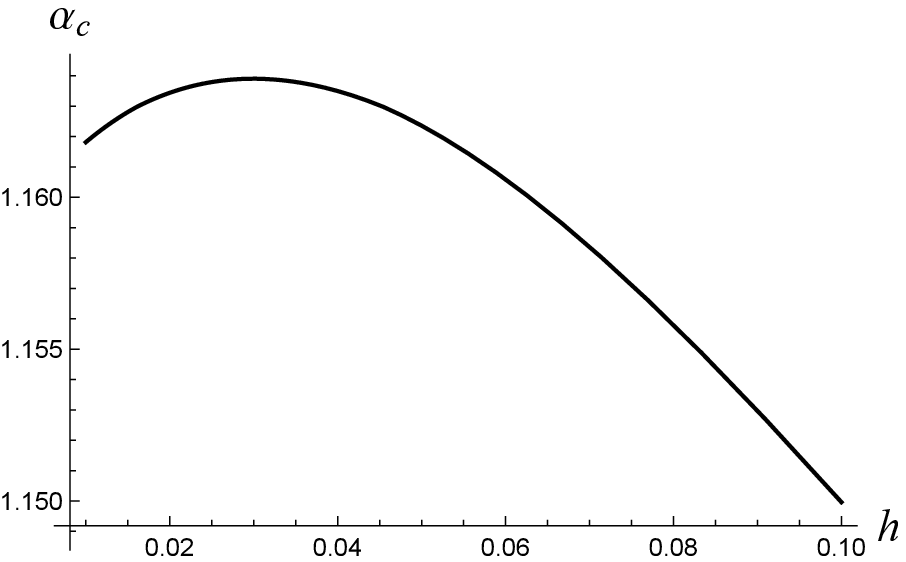}{\label{fig4f}}}}
\caption{Dependence of the transition angle $\alpha_c$ on different parameters.}\label{fig4}
\end{figure}

Bearing in mind that there are several parameters involved. It is of fundamental interest to precisely reveal their effects on this transition angle $\alpha_c$. We then carry out a complete parametric analysis. Note that there are in total five free parameters at hand. But previous analysis manifests that a casual choice of $s$ may give rise to an irrational result. Thus, we fix $s=0.8$ in all illustrative examples. Furthermore, a two dimensional picture is obviously more clear to exhibit the result. For that purpose, we should specify three of those parameters and depict $\alpha_c$ as a function of the remaining  parameter. In doing so, an entire parametric study is shown in Figure \ref{fig4}. In particular, Figures \ref{fig4a}--\ref{fig4c} are concerned with the effects of material constants while Figures \ref{fig4d}--\ref{fig4f} focus on the dependences on geometric parameters. Figure \ref{fig4a} plots the relation between $\alpha_c$ and $\mu_f$, which is a monotonically increasing function. A dashed line (see also Figures \ref{fig4b} and \ref{fig4c}) for $\alpha_c=\pi$, which illustrates the limit of $\alpha$, is also shown for comparison. If $\mu_f$ roughly exceeds 15, the value of $\alpha_c$ will be larger than $\pi$, which is the limit value of $\alpha$, such that director reorientation becomes impossible. In Figure \ref{fig4b}, we keep $\mu_f=\kappa$ and vary them simultaneously. A similar situation is observed, but the parametric region that allows director reorientation gets much smaller.  We investigate the effect of $\kappa$ on $\alpha_c$ in Figure \ref{fig4c}. Reversely, it can be seen that the more the nematic property dominates, the earlier the director transition occurs. Furthermore, Figure \ref{fig4d} directs the relevance of $\alpha_c$ to the whole structure thickness when $h=H$. The dependence of $\alpha_c$ on $H$ and the counterpart for varied $h$ are depicted in Figures \ref{fig4e} and \ref{fig4f}, respectively. It can be seen that the curves in Figures \ref{fig4d} and \ref{fig4e} are monotonically decreasing functions, which indicates that a thicker substrate becomes more likely to create a director transition. Notwithstanding, a non-monotonic curve for the relation between the transition angle $\alpha_c$ versus $h$, as shown in Figure \ref{fig4f}, occurs if the thickness of substrate is fixed. Therefore, an increase of $h$ can enlarge the value of transition angle $\alpha_c$ when the LCE film is extremely thin but reduces that after it passes a critical value. Furthermore, it is seen that the maximum is near $h=0.03$ in this case. 

According to the above observations, it follows that the solution $\beta=\dfrac{\pi}{2}$ is always stable at an earlier stage of deformation. Yet it may give way to another solution $\beta=0$ at a critical angle $\alpha_c$ for certain material and geometric parameters. In this case, the director rotates $\dfrac{\pi}{2}$ abruptly, and a director orientation (or transition) takes place. Moreover, a complete parametric study for $\alpha_c$ implies that there is a competition among those parameters. A stiffer NLCE tends to delay the director transition while either a stronger $\kappa$ or a thicker substrate can trigger the director transition earlier. However, the effect of the film thickness is parameter-dependent and can either be supportive of or deterrent to a director reorientation.

\section{Concluding remarks}
We have solved the finite bending of an NLCE-substrate structure exactly in the frame work of nonlinear elasticity. In particular, the NLCE film is modeled by a two-constant strain-energy function proposed by \cite{fried1999}, and the substrate is composed of an incompressible neo-Hookean material. The reduced governing system together with the boundary conditions and continuity condition were formulated and solved exactly. Interestingly, a pair of non-trivial solutions was found with each solution having its own director. Furthermore, the two directors are mutually perpendicular. We then addressed the issue of the preferred solution by comparing the total potential energy. It is found that, for some parameters, the director $\bm n$ rotates suddenly from $\bm e_r$ to $\bm e_{\theta}$ at a critical bending angle $\alpha_c$ which is referred to the transition angle. We identify this as a director reorientation, which has been observed in uni-axial extension experiments \citep{prl1993}. We mention that finite bending of NLCEs has been addressed by \cite{pence2006} using a different constitutive model. However, such a director reorientation was not addressed. Based on the exact solutions, a detailed parametric study was conducted which precisely reveal the influences of different material and geometric parameters on the transition angle $\alpha_c$. Meanwhile, the obtained exact solution can also be used as a benchmark problem to validate the accuracy of different plate models for NLCEs.

\section*{Acknowledgments}
The work described in this paper was supported by a grant from the National Natural Science Foundation of China (Project No: 11602163).

\end{document}